\documentclass[prd,aps,showpacs,secnumarabic,superscriptaddress,twocolumn]{revtex4}
\usepackage{graphicx}
\usepackage{bm}
\usepackage{amsmath}
\usepackage{amsfonts}
\usepackage{amssymb}
\def\Dot{\!\cdot\!}
\def\Trip{$^3\!P_s\;$}
\def\Sing{$^1\!P_s\;$}

\def\ep{\varepsilon}
\begin{document}
\title{Noncommutative QED and the Lifetimes of Ortho and Para Positronium  }
\author{Mauro Caravati}
\email{Mauro.Caravati@ca.infn.it}
\affiliation{Dipartmento di Fisica, Universit\'a di Cagliari and INFN, Sezione di Cagliari, Cagliari, Italy}
\author{Alberto Devoto}
\email{Alberto.Devoto@ca.infn.it}
\affiliation{Dipartmento di Fisica, Universit\'a di Cagliari and INFN, Sezione di Cagliari, Cagliari, Italy} 
\author{Wayne W. Repko}
\email{repko@pa.msu.edu} 
\affiliation{Department of Physics and Astronomy, Michigan State University, East Lansing, MI 48824}

\date{\today}

\begin{abstract}
We examine the implications of including corrections associated with the noncommutative extension of quantum electrodynamics - NCQED - to the decays of the ortho and para positronium (\Trip and \Sing) ground states. In NCQED, the well known charge conjugation argument restricting \Trip decays to an odd number of photons and \Sing decays to an even number of photons no longer obtains. Instead, the dominant two photon decay mode of \Sing is accompanied by a three photon mode. The dominant three photon decay mode of \Trip receives a NCQED correction, but there is no corresponding two photon decay mode in the weak binding limit. These corrections to the $P_s$ three photon decay mode have a different energy distribution, but their effect is too small to explain any discrepancy between the observed and calculated values of the \Trip lifetime. 
\end{abstract}
\pacs{11.10.Nx,36.10.Dr}
\maketitle

\section{Introduction}

\begin{figure}[h]
\hfill\includegraphics[height=.5in]{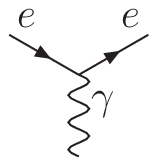}
\hfill\includegraphics[height=.55in]{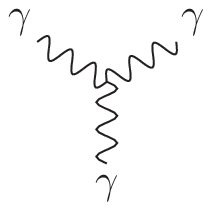}
\hfill\includegraphics[height=.45in]{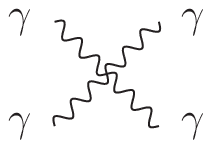}
\hfill\includegraphics[height=.5in]{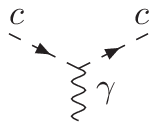}\hspace*{0pt\hfill}
\caption{\footnotesize The basic vertices of NCQED. $c$ denotes the ghost.\label{ncqedvert}}
\end{figure}

The interest in formulating field theories on noncommutative spaces is relatively old \cite{snyder}. It has been revived recently due to developments connected to the analysis of string theories \cite{cds,dh,sw}. These developments suggest that field theories on non-commuting spaces are well defined quantum theories \cite{dh}. In noncommutative geometry, the coordinates $x^{\mu}$ obey the commutation relations
\begin{equation}
\left[x^{\mu},x^{\nu}\right]= i\theta^{\mu\nu}\,,
\end{equation}
where $\theta^{\mu\nu}=-\theta^{\nu\mu}$. A noncommutative version of an ordinary field theory can be obtained by replacing all ordinary products with Moyal $\star$ products defined by
\begin{equation}
(f\star g)(x)=\left.\exp\left(\textstyle\frac{1}{2}\theta^{\mu\nu}\partial_{x^{\mu}}
\partial_{y^{\nu}}\right)f(x)g(y)\right|_{x=y}\,.
\end{equation}
In particular, the application of this recipe to the Lagrangian for ordinary quantum electrodynamics (QED) results in the noncommutative quantum electrodynamics (NCQED) Lagrangian
\begin{equation}\label{lag}
\mathcal{L}=\textstyle\frac{1}{2}i\bar{\psi}\star\gamma^{\mu}\stackrel{\leftrightarrow}{D_{\mu}}
\psi - m\bar{\psi}\star\psi - \textstyle\frac{1}{4}F_{\mu\nu}\star F^{\mu\nu}\,,
\end{equation}
where
\begin{eqnarray}
D_{\mu}\psi & = & \partial_{\mu}\psi - ieA_{\mu}\star\psi\,, \\
D_{\mu}\bar{\psi} & = & \partial_{\mu}\bar{\psi}+ie\bar{\psi}\star A_{\mu}\,, \\
\bar{\psi}\star\stackrel{\leftrightarrow}{D_{\mu}}\psi & = & \bar{\psi}\star D_{\mu}\psi - D_{\mu}\bar{\psi}\star\psi\,, \\[4pt]
F_{\mu\nu}& = & \partial_{\mu}A_{\nu}-\partial_{\nu}A_{\mu}-ie\left[A_{\mu},A_{\nu}\right]_{\star}\,,
\end{eqnarray}
and
\begin{equation}
\left[A_{\mu},A_{\nu}\right]_{\star}=A_{\mu}\star A_{\nu}-A_{\nu}\star A_{\mu}\,.
\end{equation}
When supplemented with a gauge fixing term, including a ghost contribution, the Lagrangian Eq.\,(\ref{lag}) can be used to obtain a set of vertices and Feynman rules for perturbative calculations \cite{rsj}. The vertices of NCQED, which include nonabelian interactions among photons, are illustrated in Fig.\,\ref{ncqedvert}. 

The physical implications of these nonabelian photon interactions have been the subject of numerous investigations \cite{hk1} including collider signatures \cite{hpr,maha,math}, effects in atomic systems \cite{csjt,lane,MPR}, contributions to the muon anomalous magnetic moment \cite{rsj,haya}, and violations of discrete symmetries \cite{sj,hk,gl}.    

\section{NCQED Corrections to the Ortho and Para Positronium Lifetimes}

In ordinary QED, difference in the lifetimes of \Sing and \Trip is usually explained using charge conjugation symmetry; \Sing is $C$ even and able to decay into two photons while \Trip is $C$ odd and obliged to decay into at least three photons. Furthermore, $C$ invariance forbids the decay $^1P_s\to 3\gamma$. The noncommutative extension of QED, NCQED, introduces photon self interactions into the theory, and the $C$ invariance arguments no longer obtain. As in case of pion decay \cite{dicus}, treated using NCQED in Ref.\,\cite{gl}, the non-Abelian photon interactions lead to a three photon decay mode for \Sing. The existence of photon self interactions does not lead to a two photon decay mode of \Trip in the weak binding approximation. There are, however, NCQED corrections to the three photon decay of \Trip. 

Since positronium is weakly bound, it is possible to calculate the NCQED correction to the lifetimes by computing the annihilation amplitudes for a free electron and positron at rest and supplying a factor of the square of the bound state wave function at the origin, $|\psi(0)|^2$, to account for the leading binding effect. There is no need to devise an effective interaction as in the case of the pion decay into three photons \cite{gl}. The NCQED amplitudes contributing to the three photon corrections were calculated using the Feynman rules of Ref.\,\cite{rsj}. These rules contain contributions involving $\theta_{\mu\nu}$ of the form $k_1^{\mu}\theta_{\mu\nu}k_2^{\nu}$, where $k_1$ and $k_2$ are the momenta of two of the photons. To ensure that the unitarity conditions $\theta_{\mu\nu}\theta^{\mu\nu}>0$ and $\ep_{\mu\nu\lambda\rho}\theta^{\mu\nu}\theta^{\lambda\rho}=0$ are satisfied, we take $\theta_{0k}=-\theta_{k0}=0$ and write $\theta_{ij}$ as
\begin{equation}
\theta_{ij}=\frac{1}{\Lambda^2}\varepsilon_{ijk}\theta_k,
\end{equation}
where $\theta_k$ is a unit vector and $\Lambda$ is the noncommutativity scale. We then have $k_1^{\mu}\theta_{\mu\nu}k_2^{\nu}=\theta\Dot(\vec{k}_1\times\vec{k}_2)/\Lambda^2$.

\begin{figure}[h]
\hfill\includegraphics[height=1.0in]{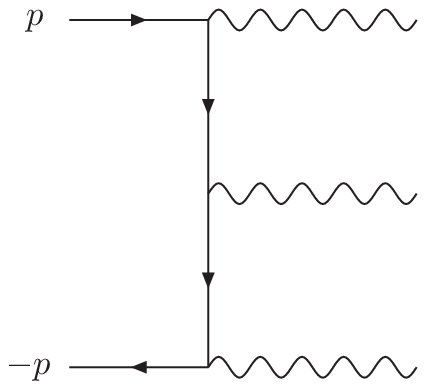}
\hfill\includegraphics[height=1.0in]{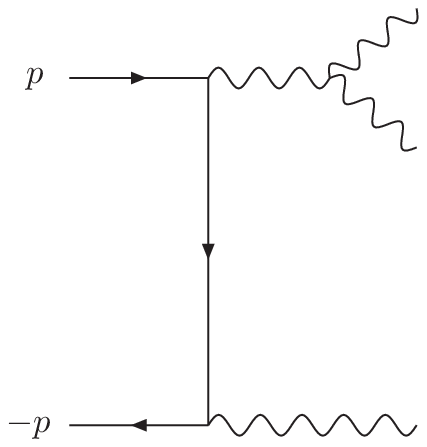}\hspace*{0pt\hfill}
\caption{\footnotesize The left diagram is illustrative of the class of {\em abelian} contributions to the positronium three photon decays and the right diagram is one of the class of {\em nonabelian} contributions. \label{diags}}
\end{figure}

The amplitudes to be calculated are illustrated in Fig.\,\ref{diags}. In the threshold approximation tree diagrams  involving four photons vanish. 
These amplitudes were calculated with the aid of the symbolic manipulation program FORM \cite{form} and it is convenient to use the metric in which the mass shell condition is $p^2=-m^2$. As a partial check of our calculation, we verified that amplitudes satisfy the appropriate Ward identities under gauge replacements $\ep^{(i)}\to k_i$, where $\ep^{(i)}$ is the polarization vector corresponding to the photon with momentum $k_i$. The sum over photon polarizations in the squared amplitude was performed in two ways. In the first, we used the usual relation
\begin{equation}\label{polar}
\sum_{\rm pol}\ep^*_{\mu}\ep_{\nu}=\delta_{\mu\nu}
\end{equation}
and subtracted the ghost contribution, while in the second we used \cite{gl}
\begin{equation}
\sum_{\rm pol}\ep^{(i)*}_{\mu}\ep^{(i)}_{\nu}=\delta_{\mu\nu}+ \frac {k_{i \mu} k_{j \nu} + k_{i \nu} k_{j \mu} }{k_i \cdot k_j} \quad 
j \neq i\,.
\end{equation}
The results agree and provide another check on the calculation. We present our results for the squared amplitude as four separate contributions: $|\mathcal{M}_A|^2$, the contribution from the {\em abelian} diagrams; $\mathcal{M}_A\mathcal{M}^*_{N}+\mathcal{M}^*_A\mathcal{M}_N$, the interference term between the {\em abelian} and {\em nonabelian} diagrams, $|\mathcal{M}_N|^2$, the contribution from the {\em nonabelian} diagrams; and $|\mathcal{M}_G|^2$, the ghost contribution. To simplify the resulting expressions, we introduce the definitions
\begin{eqnarray} \label{defs}
z_1 & = & p\Dot k_1, \qquad z_2 = p\Dot k_2, \qquad z_3 = p\Dot k_3\,, \\
y_1 & = & k_2 \Dot k_3, \hspace*{16pt} y_2 = k_3 \Dot k_1, \hspace*{15pt} y_3 = k_1 \Dot k_2\,, \\
F & = & z_1 z_2 + z_2 z_3 + z_3 z_1\,.
\end{eqnarray}
These variables are not all independent. They are constrained by the threshold energy-momentum conservation relation $2p=k_1+k_2+k_3$, where, at threshold, $\bar{p}_{\mu}=p_{\mu}=(0,0,0,im)$. In terms of these variables, the four contributions to the squared matrix element summed over photon spins using Eq.\,(\ref{polar}) are
\begin{eqnarray}
|\mathcal{M}_A|^2  = &4e^6&\hspace{-4pt}\frac{\sin^2 \varphi}{z_1 z_2 z_3}\left[ \frac {8 F} {p^2}
- \frac {p^2 \left( z_1^3 + z_2^3 +z_3^3\right)}{z_1z_2z_3}\right. \nonumber \\ 
& + &\left.\frac{p^2\left(13 z_1 z_2 z_3 - 2p^2 F \right)} {z_1 z_2 z_3} \right]\,,
\end{eqnarray}
\begin{eqnarray}
|\mathcal{M}_N|^2= 4e^6\,\frac{\sin^2 \varphi}{p^2} \nonumber \\
& &\hspace*{-36pt}\times\left[\frac {8 y_1 y_2 y_3 + 2 z_2\left(3 y_2^2-y_1 y_3 \right)}{y_1y_3z_1z_3}\right.\nonumber \\ 
& &\hspace*{-30pt}+\left.\frac{y_2 \left(y_1^2+y_3^2+2y_2^2\right)}{y_1 y_3 z_1 z_3} \right. \nonumber \\
& &\hspace*{-30pt}+\left.\frac {8 y_1 y_2 y_3 + 2 z_3\left(3 y_3^2-y_1 y_2 \right)}{y_1 y_2 z_1 z_2}\right. \nonumber \\ 
& &\hspace*{-30pt}+\left.\frac{y_3 \left(y_1^2+y_2^2+2y_3^2\right)}{y_1 y_2 z_1 z_2}\right. \nonumber \\
& &\hspace*{-30pt}+\left.\frac {8 y_1 y_2 y_3 + 2 z_1\left(3 y_1^2-y_2 y_3 \right)}{y_2 y_3 z_2 z_3}\right. \nonumber \\ 
& &\hspace*{-30pt}+\left.\frac{y_1 \left(y_2^2+y_3^2+2y_1^2\right)}{y_2 y_3 z_2 z_3}\right. \nonumber \\
& &\hspace*{-30pt}-\left.\frac {3 y_2 y_3 +4 \left(y_2^2+y_3^2 \right)} {2 y_1 z_1^2}\right.\nonumber \\ 
& &\hspace*{-30pt}-\left.\frac {3 y_1 y_3 +4 \left(y_1^2+y_3^2 \right)} {2 y_2 z_2^2}\right. \nonumber \\ 
& &\hspace*{-30pt}\left.\frac {3 y_1 y_2 +4 \left(y_1^2+y_2^2 \right)} {2 y_3 z_3^2} \right]\,,
\end{eqnarray}
\begin{eqnarray} 
\mathcal{M}_A\mathcal{M}^*_N&+&\mathcal{M}^*_A\mathcal{M}_N=4e^6\,\frac {\sin^2 \varphi}{p^2}\nonumber \\ 
& &\hspace*{-36pt}\times\left[\frac {5 y_3 \left( y_1-y_2 \right) + 2 z_3 \left( 2y_1 + y_2 \right)}{2 z_1 z_3 y_3} + \frac {y_2^2} {z_1 z_3^2}\right. \nonumber \\
& &\hspace*{-30pt}\left.+ \frac {5 y_1 \left( y_2-y_3 \right) + 2 z_1 \left( 2y_2 + y_3 \right)}{2 z_1 z_2 y_1} + \frac {y_3^2} {z_1^2 z_2}\right.\nonumber \\
& &\hspace*{-30pt}\left.+ \frac {5 y_2 \left( y_3-y_1 \right) + 2 z_2 \left( 2y_3 + y_1 \right)}{2 z_2 z_3 y_2} + \frac {y_1^2} {z_2^2 z_3}\right.\nonumber \\
& &\hspace*{-30pt}\left.+ \frac {5 y_1 \left( y_3-y_2 \right) + 2 z_1 \left( 2y_3 + y_2 \right)}{2 z_1 z_3 y_1} + \frac {y_2^2} {z_1^2 z_3}\right.\nonumber \\
& &\hspace*{-30pt}\left.+ \frac {5 y_2 \left( y_1-y_3 \right) + 2 z_2 \left( 2y_1 + y_3 \right)}{2 z_1 z_2 y_2} + \frac {y_3^2} {z_1 z_2^2}\right.\nonumber \\
& &\hspace*{-30pt}\left.+ \frac {5 y_3 \left( y_2-y_1 \right) + 2 z_3 \left( 2y_2 + y_1 \right)}{2 z_2 z_3 y_3} + \frac {y_1^2} {z_2 z_3^2}\right.\nonumber \\
& &\hspace*{-30pt}\left.-\frac {2 y_1 y_2 y_3 - y_2 y_3 z_1 + 4 z_1^3} {z_1 z_2 z_3 y_1}\right.\nonumber \\
& &\hspace*{-30pt}\left.-\frac {2 y_1 y_2 y_3 - y_1 y_3 z_2 + 4 z_2^3} {z_1 z_2 z_3 y_2}
\right.\nonumber \\
& &\hspace*{-30pt}\left. 
-\frac {2 y_1 y_2 y_3 - y_1 y_2 z_3 + 4 z_3^3} {z_1 z_2 z_3 y_3}
       \right]\,,
\end{eqnarray}
\begin{equation}
|\mathcal{M}_{G}|^2 = -6e^6\,\frac {\sin^2 \varphi}{p^2}
\left( \frac{y_2y_3} {y_1 z_1^2} + \frac{y_1y_3} {y_2 z_2^2}
+ \frac{y_1y_2} {y_3 z_3^2} \right)\,.
\end{equation}
These forms of the various contributions are not unique due to the relations
\begin{eqnarray}
y_1+y_2+y_3 & = & -2m^2\,, \\
z_1+z_2+z_3 & = & -2m^2\,, \\
z_1 & = & \textstyle\frac{1}{2}(y_2+y_3)\,,\\
z_2 & = & \textstyle\frac{1}{2}(y_3+y_1)\,,\\
z_3 & = & \textstyle\frac{1}{2}(y_1+y_2)\,.\\
\end{eqnarray}
We have attempted to present the results in a way which reveals the symmetry in the labels $1,2,3$. The factor $\sin\varphi$ is
\begin{equation}\label{sinphi}
\sin\varphi=\sin\left(\frac{\vec{\theta}\Dot(\vec{k}_1\times\vec{k}_2)}{2\Lambda^2}\right)\simeq
\frac{\cos\delta\,\omega_1\omega_2\sin(\theta_{12})}{2\Lambda^2}\,,
\end{equation}
where $\delta$ is the angle between $\vec{\theta}$ and the normal to the plane of the vectors $\vec{k}_1,\vec{k}_2,\vec{k}_3$ and $\theta_{12}$ is the angle between $\vec{k}_1$ and $\vec{k}_2$. The integration over phase space in the positronium rest frame \cite{gl} with the inclusion a minus sign for the ghost contribution and factor of $1/3!$ for identical particles gives 
\begin{eqnarray}
\mathcal{I}_A &=& \alpha^3\frac {a^2} {m^2} 
\left( \frac {19 \pi^2}{3} - \frac {562} {9}\right)\,,\\
\mathcal{I}_{N} &=& \alpha^3\frac {a^2} {m^2} 
\left(\frac {16 \pi^2}{3} - \frac {623} {12}\right)\,,\\
\mathcal{I}_{AN} &=& \alpha^3\frac {a^2} {m^2} 
\left(\frac {2 \pi^2}{3} - \frac {58} {9}\right)\,,\\
\mathcal{I}_G &=& \alpha^3\frac {a^2} {m^2} \left(\frac {1} {120}\right)\,,
\end{eqnarray}
where
\begin{equation}
a^2 = \left(\frac{m^2}{\Lambda^2}\right)^2 \cos^2\delta\,.
\end{equation}
Including the factor $|\psi(0)|^2$, the decay width of \Sing into three photons arising from NCQED is \cite{gl}
\begin{equation}
\frac{d\Gamma_{^1\!P_s \to 3\gamma}}{d\cos\delta} =
\frac {m \alpha^6} {24 \pi}\left(\frac{m^2}{\Lambda^2}\right)^2\!\cos^2\!\delta \left(37 \pi^2 - \frac {5434}{15} \right). 
\end{equation}

For the \Trip case, the calculation is straightforward because the only contribution comes from the {\em abelian} diagrams of Fig.\,\ref{diags} and one finds the standard QED amplitude \cite{op} $^3\mathcal{M}$ multiplied by an overall $\cos \varphi$ factor. The squared amplitude is
\begin{eqnarray}
|^3\mathcal{M}|^2 &=& \frac {16}{3} e^6\cos^2\varphi\left[\frac{\left(z_1-p^2 \right)^2}{z_2^2z_3^2}\right.\nonumber \\
& &\left. + \frac{\left(z_2-p^2 \right)^2}{z_1^2z_3^2} 
     + \frac{\left(z_3-p^2 \right)^2}{z_1^2 z_2^2}\right].
\end{eqnarray}
Expanding the cosine and integrating over phase space we find that
the NCQED correction to the three photon decay of \Trip is
\begin{equation}
\frac{d\Gamma^{(NC)}_{^3\!P_s \to 3\gamma}}{d\cos\delta} =
\frac {m \alpha^6}{24 \pi}\left(\frac{m^2}{\Lambda^2}\right)^2\!\cos^2\!\delta \left(\frac {52}{3}\pi^2 - \frac {1540} {9} \right).
\end{equation}

\section{Discussion and Conclusions}

\begin{figure}[h]
\vspace{8pt}
\centering
\includegraphics[height=2.2in]{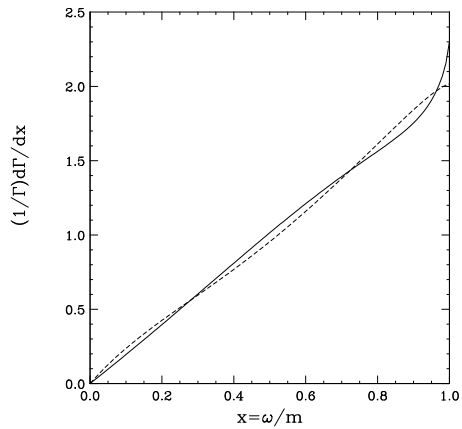}
\caption{The photon energy distributions $(1/\Gamma)d\Gamma/dx$, where $x=\omega/m$ are shown for the three photon decays of \Trip (solid) and \Sing (dashed). \label{dGdx}}
\end{figure}

The energy distribution of the photons from the three photon decay of \Sing,
\begin{equation}
\frac{1}{\Gamma_{^1\!P_s \to 3\gamma}}\frac{d\Gamma_{^1\!P_s \to 3\gamma }}{dx}\quad x=\omega/m\,,
\end{equation}
is compared with that of the photons from the three photon decay of \Trip in Fig.\,\ref{dGdx}.
In principle, it is possible to distinguish the three photon \Sing decays from those of \Trip, but, with any sensible scale for the noncommutativity \cite{ABDG,CCL}, the number of \Sing decays is too small to account for a discrepancy between the calculated and observed \Trip lifetime \cite{afs}. Having said this, it is worth noting that, due to the factor $\sin\varphi$, Eq.\,(\ref{sinphi}), in the three photon coupling, the NCQED contributions to the three photon decays do not introduce any infrared divergences. In ordinary QED, the absence of an infrared divergence in the \Trip$\to 3\gamma$ photon spectrum can be understood (at the one loop level, at least) in the following way. The usual method of removing a soft photon singularity by combining it with the one loop correction to the same process with one less photon is not available in \Trip decay because \Sing$\to 2\gamma$, the process with one less photon, has the opposite $C$. This being the case \Trip$\to 3\gamma$ must be infrared finite. The same argument accounts for the fact that the one loop QED corrections to \Sing$\to 2\gamma$ are infrared finite. The introduction of NCQED nonabelian interactions could have disrupted this situation insofar as \Sing$\to 3\gamma$ is concerned because there are additional one loop corrections to \Sing$\to 2\gamma$ which are available to cancel potential infrared singularities and $C$ invariance no longer prohibits this. 
\acknowledgements
We would like to thank Yi Liao for helpful comments. One of us (WWR) wishes to thank INFN, Sezione di Cagliari and Dipartmento di Fisica, Universit\'a di Cagliari for support. This research was supported in part by the National Science Foundation under Grant PHY-0070443 and by M.I.U.R. (Ministero dell'Istruzione, dell'Universit\`a e della
Ricerca) under Cofinanziamento PRIN 2001. .

\end{document}